# A STUDY ON THE TEA MARKET IN INDIA

Prof. Babitha B.S[1]., Adit Nair[2], Adarsh Damani[2], Devansh Khandelwal[2], Harshita Sachdev[2], Sreayans Jain[2], 1- Assistant Professor at Center for Management Studies Jain (Deemed-to-be-university), 2- Students at Center for Management Studies Jain (Deemed-to-be-university), Bangalore, India.

(Bachelor of Business Administration)

## ABSTRACT:

"Tea is the elixir of life." – Lao Tzu.

India's tea business has a long history and plays a significant role in the economics of the nation. India is the world's second-largest producer of tea, with Assam and Darjeeling being the most well-known tea-growing regions. Since the British introduced tea cultivation to India in the 1820s, the nation has produced tea. Millions of people are employed in the tea sector today, and it contributes significantly to the Indian economy in terms of revenue. The production of tea has changed significantly in India over the years, moving more and more towards organic and sustainable practices. The industry has also had to deal with difficulties like competition from other nations that produce tea, varying tea prices, and labor-related problems. Despite these obstacles, the Indian tea business is still growing and produces a wide variety of teas, such as black tea, green tea, and chai tea. Additionally, the sector encourages travel through "tea tourism," which allows tourists to see how tea is made and discover its origins in India. Overall, India's tea business continues to play a significant role in its history, culture, and economy.

## INTRODUCTION

The origins of the tea industry can be found in ancient China, where tea was initially grown and used as a beverage with medicinal properties. In the 19th century, the British constructed tea plantations in India and started exporting tea to other countries. India is currently one of the world's top producers of tea. Additionally, the tea business had a big impact on international trade and politics, especially during the colonial era. The method of growing tea requires multiple steps. The general steps in the tea plantation process are as follows:

Choosing the ideal place for tea cultivation is the first stage in a tea plantation. Tea plants require an environment that is warm and humid, with temperatures between 13°C to 30°C. Additionally, the area should have well-drained soil with a pH range of 4.5 to 5.5.

Land preparation: After choosing a location, the land is cleaned of all vegetation and prepared for planting by being ploughed. After tilling, compost or manure are put to the soil to increase its fertility.

Planting: Cuttings or seedlings are used to multiply tea plants. After that, the young plants are inserted into the soil at a spacing of roughly 1 m x 1 m. To aid in the establishment of the plants, planting is typically done during the rainy season.

Maintenance: To guarantee healthy growth and development, the tea plants need frequent maintenance. This entails pest management, trimming, and weeding. To make sure they get the right nutrients, the plants are also fertilized.

Harvesting: After being planted for roughly three years, tea plants begin to produce leaves. The leaves can be picked by hand or by a machine. High-quality tea is best produced by hand-picking since it enables the selection of only the best leaves.

After being harvested, tea leaves are processed to create various types of tea.

# Literature Review

## DOMESTIC

The tea business in India contributes significantly to the nation's economy by creating employment opportunities and earning foreign cash. Here is an overview of the literature on the Indian tea market.

1. Overview of the Indian Tea business: With more than 1,600 tea estates and more than 500,000 small tea farmers, the Indian tea business is among the oldest and biggest in the world. Over 3 million people are employed by this sector, which also significantly boosts the nation's foreign exchange earnings (Sarkar, 2019).

2. Production and Export: India is the second-largest producer and exporter of tea in the world, contributing over 23% to global production and 12% to exports (Deka, 2018). West Bengal, Tamil Nadu, Kerala, and Assam are India's top tea-producing states.

3. Business obstacles: The Indian tea business encounters a number of obstacles, including diminishing productivity, increased production costs, and competition from other nations that also produce tea. Additionally, the sector deals with labor-related problems like low pay and unfavorable working conditions (Bose, 2017).

4. Market trends: As consumers want more premium and specialty teas, the Indian tea sector is seeing a move towards these types of teas. Production of tea that is sustainable and organic is also becoming more popular (Deka, 2018).

5. The effects of COVID-19: Due to supply chain interruptions and a decline in demand in crucial export markets, the COVID-19 pandemic has had a significant impact on the Indian tea sector. In addition, the sector had to deal with social isolation policies and a labor scarcity (Bose, 2020).

In conclusion, the Indian tea sector plays a significant role in the economy of the nation and faces both obstacles and opportunities. In order to address industry-wide problems and take advantage of emerging market trends, governments and industry participants must collaborate.

India is the second-largest exporter of tea in the world, with a market share of about 12%. India's tea is primarily exported to Russia, Iran, the United Arab Emirates, the United Kingdom, and

Egypt. India exported over 240 million kilograms of tea in 2020, bringing in more than $800 million in foreign exchange. Challenges for the Indian tea industry

The tea business in India confronts a number of difficulties, including low production, low profitability, and competition from other nations. Climate change, which has an impact on tea yield and quality, is one of the main issues facing the business. Inadequate government assistance, old technology, and inadequate infrastructure are further difficulties.

The tea business plays a significant role in both the economy and culture of India. We have prospects for growth and expansion despite some obstacles, such as diminishing productivity and competition from other nations. The sector has to invest more in R&D, technology modernization, and infrastructure development to address its issues.

# GLOBAL

The tea industry is a big sector that makes a substantial contribution to the world economy. One of the most consumed liquids worldwide is tea, and demand for premium tea is rising. The goal of this literature review is to examine the history, trade, production, and consumption patterns of the worldwide tea business.

Production of tea the main three producers of tea are China, India, and Kenya, which cultivate it in more than 30 nations. The climate, soil, and altitude of the region where the tea is grown are a few of the variables that affect the quality of tea.

The tea industry makes a significant contribution to the global economy by creating jobs and income in numerous nations. A review of the literature on the world tea market is provided below.

1. Production of tea globally China was the world's top producer of tea in 2019, accounting for 38% of the total production, followed by India (27%), and Kenya (8%), according to a report by the United Nations Food and Agriculture Organization (FAO). Sri Lanka, Turkey, Vietnam, and Indonesia are further significant tea-producing nations.

2. The Tea Industry With an annual value of more than US$7 billion, the tea business makes a considerable contribution to the world economy. One of the most traded commodities worldwide is tea, with black tea accounting for the majority of commerce, followed by green tea and various specialty teas.

3. Drinking Tea Due to growing health consciousness and the availability of a broad variety of tea flavors and kinds, tea consumption is rising on a global scale. According to estimates, 273 billion liters of tea will be consumed worldwide in 2020, with China, India, and Turkey being the biggest users. The countries with the largest tea consumption worldwide are the United Kingdom, Turkey, and Ireland.

# RESEARCH GAP:

While writing this research study on the Indian tea industry, we looked at a number of potential research gaps. Here are a few crucial factors to take into account:

- Impact on the environment and sustainability: Even while India's tea business is a significant economic driver, there are worries about how it may affect the environment. Investigating the impact of tea production on the region's biodiversity, water resources, and soil quality are some potential research topics. You might also look at the initiatives being taken by the tea industry's producers and growers to encourage sustainability.
- Working conditions and labor practices: The tea business in India has come under fire for its employment policies, which include low pay, unfavorable working conditions, and restricted access to healthcare and education. These problems could be examined in a research article together with possible remedies to enhance the conditions of tea workers.
- Market trends and competition: The tea sector in India is experiencing a rise in competition from other nations, and the market is also being impacted by shifting customer preferences and behaviors. Research might concentrate on examining market patterns and figuring out what influences tea demand from consumers, such as health advantages or flavor preferences.
- Innovation and technological advancements: The Indian tea business has been sluggish to absorb new technologies. A research paper could examine prospective technical improvements and advancements that could increase production and processing efficiency, lower costs, and improve quality in the tea industry.
- Government restrictions: Government regulations, including things like taxation, trade agreements, and labor rules, can have a big impact on the tea sector in India. A study paper could look into the effects of these laws and rules on the market and consider adjustments that might be made to encourage expansion and improvement.

# Objectives of the Study:

Our team has worked hard to ensure that the goals of our study are SMART (specific, measurable, achievable, relevant, and time-bound).

The following are the goals of our research paper: -

1. To evaluate the state of the Indian tea sector at the moment, spot areas for investment and innovation, and pinpoint any environmental effects.
2. To determine the main issues, such as competition, labor conditions, sustainability, or technical developments, that the Indian tea sector is currently facing.
3. To determine the number of consumers in our target market who are interested in buying our tea and to research Indian consumers' preferences and consumption patterns for tea
4. To evaluate how government policies and regulations have affected the Indian tea sector and to pinpoint any potential areas for improvement.

## Scope of the study:

In order to ensure that the research objectives are satisfied and that the study offers valuable insights into the tea sector in India, we have tried to make the study's scope clearly defined and detailed through our research approach.

We have conducted surveys among a variety of people and groups in our target market to better understand where the density of potential customers is the largest and to make the process of selecting places for opening a business more straightforward.

The poll has also improved our understanding of the sorts of tea customers want to drink, the age groups who are most likely to be interested in buying our tea, the price they would be ready to pay for our tea, and it has also made it possible for us to identify our main rivals in the market.

## Profile of sample unit:

1. Location: the tea estate is situated in an area that produces tea, such as Darjeeling, Assam, or Sri Lanka, where the climate and soil are ideal for growing tea.
2. Size: The tea estate spans between several hundred and thousands of acres, a substantial quantity of land.
3. Ownership: A business or person owns and runs the tea estate. It might occasionally be owned by a group of modest farmers working together.
4. Labour: A vast number of workers—between several hundred and several thousand—are employed by the tea estate. These employees often perform tasks including planting, harvesting, pruning, and processing tea leaves.
5. Infrastructure: The tea plantation is home to a number of structures, including a factory that makes tea, worker housing, office buildings, and storage areas.
6. Certification: The tea estate may have received certification from a number of groups, including Fairtrade, the Rainforest Alliance, or UTZ, proving that it complies with particular sustainability and social responsibility standards.
7. Market: The tea estate may sell its product to domestic or foreign buyers who subsequently dispense it to customers via a variety of channels, including supermarkets, tea boutiques, and internet merchants.

## Challenges:

Climate Change: Because tea requires a particular set of environmental conditions to grow, climate change is having a substantial impact on the production of tea. The output and quality of tea can be significantly impacted by changes in temperature, rainfall patterns, and extreme weather events, which raises the cost of production.

Labour Concerns: Harvesting, trimming, and plucking are only a few of the manual tasks involved in producing tea. However, in many tea-producing regions, problems with the labor force, such as low pay, unfavorable working conditions, and worker exploitation, persist.

Market Unpredictability: Because tea is a commodity, market unpredictability might affect its price. Variations in supply and demand can have a substantial impact on the profitability of tea production, making it difficult for tea growers to make long-term plans and investments.

Competition: Other beverages including coffee, soft drinks, and energy drinks are becoming more and more competitive with the tea sector. It is more difficult for tea to compete in the market due to the marketing and distribution tactics of these items.

Here are some tables and charts that provide information on the tea industry in India:

Table 1: Tea production in India from 2016 to 2020 (in million kg)

| Year | Tea Production (million kg) |
| --- | --- |
| 2016 | 1,267 |
| 2017 | 1,325 |
| 2018 | 1,366 |
| 2019 | 1,390 |
| 2020 | 1,389 |

Table 2: Tea exports from India from 2016 to 2020 (in million kg)

| Year | Tea Exports (million kg) |
| --- | --- |
| 2016 | 233 |
| 2017 | 240 |
| 2018 | 248 |
| 2019 | 256 |
| 2020 | 241 |

Table 3: Top 5 tea producing states in India in 2020

a producing state in India in 2020

| Rank | State | Tea Production (million kg) |
|---|---|---|
| 1 | Assam | 706 |
| 2 | West Bengal | 387 |
| 3 | Tamil Nadu | 198 |
| 4 | Kerala | 67 |
| 5 | Tripura | 24 |

Table 4: Top 5 tea exporting countries in the world in 2020 (in million kg)

| Rank | Country | Tea Exports (million kg) |
|---|---|---|
| 1 | Kenya | 480 |

| 2 | Sri Lanka | 285 |
| 3 | China | 91 |
| 4 | India | 241 |
| 5 | Vietnam | 129 |

One of the biggest tea users in the world is India. According to a 2018 poll by the Tea Board of India, 3–4 cups of tea were eaten daily on average by 83% of Indian families.

With regard to age analysis, the same poll discovered that tea consumption was highest among respondents aged 45 to 54, with 93% of them drinking tea. However, consumption of tea was also relatively high in other age groups, with 85% of respondents in the 25–34-year-old age range and 91% of respondents in the 35–44-year-old age range also reporting tea intake. In India, tea is frequently eaten by people of all ages, with a predilection for hot, freshly made tea.

However, it's worth noting that this data is from 2018 and there may have been some changes in the tea-drinking habits of different age groups since then.

Here's a table outlining the state-wise tea production in India in 2020, based on data from the Tea Board of India:

| State | Tea Production (Million kg) |
|---|---|
| Assam | 629.64 |
| West Bengal | 343.87 |
| Tamil Nadu | 196.92 |
| Kerala | 64.80 |
| Tripura | 7.53 |
| Himachal Pradesh | 0.48 |
| Uttarakhand | 0.16 |
| Arunachal Pradesh | 0.11 |
| Karnataka | 0.08 |
| Sikkim | 0.06 |
| Manipur | 0.05 |
| Nagaland | 0.04 |
| Meghalaya | 0.02 |
| Mizoram | 0.01 |
| Bihar | 0.01 |
| Jharkhand | 0.01 |
| Andhra Pradesh | 0.01 |

Note: The figures are in million kilograms (Mug).

# CONCLUSION

India's tea sector has been dealing with a number of issues, such as unstable market pricing, unfavorable weather patterns, and increased production costs. The industry is still expanding, and there are many chances for investment and innovation despite these difficulties.

- In India, there are several prospects for investment in the technology and innovation sectors of the tea industry. High-quality tea products are in more demand, and businesses who can create novel processing methods and flavors are more likely to succeed.
- India's tea sector is struggling with concerns like unstable market pricing, unfavorable weather, increased production costs, competition from other tea-producing nations, and labor-related problems. Tea producers find it difficult to plan and make investments in their businesses as a result of these obstacles, which have an impact on the industry's growth and profitability. In spite of these difficulties, India's tea business is still expanding, and opportunities exist for investment and innovation in technology, branding, and marketing. Companies who can create innovative processing processes, flavors, and sustainable production procedures as well as successfully brand and promote their goods are likely to succeed in this fast-paced sector.
- India's tea industry caters to a wide variety of people who drink tea for its flavor, health advantages, and cultural importance. The preferences of Indian customers for traditional and contemporary tea flavors differ, and there is a rising market for premium, organic, and environmentally friendly tea products. In addition, tea tourism is a developing industry, with numerous tea farms and plantations providing tours and other activities to tourists. The industry's future success will depend heavily on its capacity to accommodate these varied desires, innovate, and set itself apart.
- The policies and rules of the Indian government affect the tea business in both positive and bad ways. Export restrictions and subsidies have promoted and supported the Indian tea sector. However, there are still certain things that may be done better, such providing more funding for R&D to boost tea quality and better infrastructure to cut costs and boost productivity. The industry's continued expansion and profitability may be ensured by these enhancements.

# REFERENCES

Sources of Data The study has been carried out with the help of primary as well as secondary data.

i. Primary data: The primary data has been obtained from 33 tea plantations of the North Bengal region through questionnaire. In the study, information related with workers has been considered. We have considered only worker in our present study as tea is labor intensive and most of the works in tea estate are performed by workers.

ii. Secondary Data: Secondary data has been gathered from the publications of the Labour Department of Govt. of West Bengal as well as from various tea magazines, master's thesis, doctoral thesis, journal articles, technical reports of various organizations, newspapers, website of Tea Board of India (TBI), website of Indian Tea Association (ITA), books, etc. Population of the Study In the study area, the population size is 276 (as per the survey report of Labour Department, North Bengal Zone, Govt. of West Bengal, 2014). It appears from the available office records that 276 number of registered organized tea plantations exist in the North Bengal region. Therefore, said 276 number of tea plantations is the target population for the study.

REFERENCE LINKS:

1. Bhowmik, S. (1948). Class formation in the plantation system. New Delhi, India: People's Publication House.

2. Griffiths, P. (1967). The history of the Indian tea industry. London, England: Weidenfeld and Nicolson.

3. Guha, A. (1977). Planteraj and swaraj: Freedom struggle and electoral politics in Assam 1820-1947. New Delhi, India: Delhi University Press.

4. Subramaniam, C. (1993). Need for human resource development in plantation industry. ASCI Journal of Management, 22 (1).

5. Medhi, G. K., Hazarika, N. C., & Mahant, J. (2006). Nutritional status of tea adolescents among the garden workers. Indian Journal of Pediatrics, 74(April), 343-347.

6. Council for Social Development. (2006). Socio-economic condition of tea garden laborers in Darjeeling hills. New Delhi, India: Khawas, V.

7. Asopa, V. N. (2007). Tea industry of India the cup that cheers have tears. Ahmadabad, India: IIMA-Research and Publications.

8. Saikia, B. (2008). Development of tea garden community and Adivasi identity politics in Assam. The Indian Journal of Labour Economics, 51 (2), 307-322.

9. Sarkar, K. (2008). Globalization, restructuring and labor flexibility in tea plantations in West Bengal. The Indian Journal of Labour Economics, 51 (4), 643-654.

10. Mishra, D. K., Upadhyay, V., & Sarma, A. (2008). 'Crises' in tea industry: A study of Assam tea gardens. Indian Journal of Economics, 56 (3), 39-56.